# The upper critical field in the BiCh$_2$-based superconductors CeOBiS$_{1.7}$Se$_{0.3}$ and PrO$_{0.5}$F$_{0.5}$BiS$_{2-x}$Se$_x$ ($x$ = 0, 0.3)

Ryosuke Kiyama[1], Kazuhisa Hoshi[1], Yosuke Goto[1], Masanori Nagao[2], Yoshikazu Mizuguchi[1*]

1. Department of physics, Tokyo Metropolitan University, 1-1, Minami-osawa, Hachioji, 192-0397, Japan.
2. Center for Crystal Science and Technology, University of Yamanashi, 7-32, Miyamae, Kofu, 400-0021, Japan.

E-mail: mizugu@tmu.ac.jp

**Abstract**

We report the upper critical field ($B_{c2}$) in the BiCh$_2$-based superconductors CeOBiS$_{1.7}$Se$_{0.3}$ and PrO$_{0.5}$F$_{0.5}$BiS$_{2-x}$Se$_x$ ($x$ = 0, 0.3). Single crystals of CeOBiS$_{1.7}$Se$_{0.3}$ and PrO$_{0.5}$F$_{0.5}$BiS$_{2-x}$Se$_x$ ($x$ = 0, 0.3) were grown using the flux method. Single-crystal structural analysis revealed that the crystal structure at room temperature is tetragonal ($P4/nmm$). Through electrical resistivity and magnetization measurements, bulk superconductivity was observed in all samples. For CeOBiS$_{1.7}$Se$_{0.3}$, the in-plane $B_{c2}$ is smaller than the conventional orbital limit and Pauli limit, suggesting that ferromagnetic ordering, which has been observed in a related Ce-containing BiCh$_2$-based compound, affects $B_{c2}$ and superconductivity. In contrast, high in-plane $B_{c2}$ was observed for PrO$_{0.5}$F$_{0.5}$BiS$_{2-x}$Se$_x$ ($x$ = 0, 0.3). We propose that the in-plane $B_{c2}$ is enhanced by antisymmetric spin–orbit coupling, which arises from the lack of local inversion symmetry.

**I. Introduction**

Since the discovery of BiCh$_2$-based (Ch: S, Se) superconductors in 2012 [1-3], the BiCh$_2$-based compound family has attracted considerable attention owing to the similarities of their crystal structures with those of cuprate and FeAs-based superconductors [4-5]. A typical BiCh$_2$-based system is given by REOBiCh$_2$ (RE: rare earth), which has a layered structure composed of alternating stacks of electrically conductive BiCh$_2$ layers and insulating (blocking) REO layers. In general, the parent phase of REOBiCh$_2$ is a band insulator [3,6]. Hence, electron carrier doping to the conduction bands, which are mainly composed of Bi-$6p_x$ and Bi-$6p_y$ orbitals in the system, is required to induce superconductivity [3]. In addition, the optimization of local crystal structures is essential for the emergence of bulk superconductivity because the superconducting properties are strongly affected by the presence of in-plane local disorders (local distortion) in BiCh$_2$-based compounds [7-9]. One of the typical routes to suppress local structural disorder is to increase the in-plane chemical pressure (CP) [3,10]. There are two ways to apply a positive in-plane CP in the REOBiCh$_2$ system: (i) using a smaller RE and/or (ii) substituting Se for the S site. In particular, the flexibility of compositional manipulation of the REO blocking layers has made it possible to synthesize BiCh$_2$-based superconductors with various compositions. Although many experimental [11] and theoretical [12] investigations have been performed with various compositions, superconducting gap symmetry and the pairing glue remain unclear. Therefore, further convincing evidence and research from a new perspective are highly desired.

In 2004, the heavy fermion superconductor CePt$_3$Si without global inversion symmetry was discovered [13], which triggered various studies on noncentrosymmetric superconductivity. In non-centrosymmetric systems, antisymmetric spin–orbit coupling (ASOC) is induced, and many novel properties such as parity mixing, magnetoelectric effects [14,15], and anomalous paramagnetic pair-breaking effects [16,17] can be expected. Recently, theoretical studies have predicted that the breaking of local inversion symmetry affects the physical properties, even if the global inversion symmetry is preserved [18]. Moreover, an anomalous upper critical field ($B_{c2}$) was observed in artificial superlattices composed of the heavy-fermion superconductor CeCoIn$_5$ and normal metal YbCoIn$_5$ owing to the Rashba-type ASOC, wherein the local inversion symmetry is broken [19]. Although many systems have local inversion symmetry, spin polarization generally cancels each other. Therefore, there are only a limited number of materials in which the effect of local inversion asymmetry on the physical properties can be observed. For BiCh$_2$-based compounds, a theoretical study predicted that hidden spin polarization by local Rashba-type ASOC should exist because of site inversion asymmetry for the Bi and S sites [20]. Indeed, spin polarization attributed to the local Rashba-type ASOC was observed by spin-ARPES (SARPES) for LaO$_{0.55}$F$_{0.45}$BiS$_2$ [21]. Furthermore, notably, the type of rare earth element and Se substitution could change the amplitude of ASOC [22]. We expect that studies that include the effect of local inversion asymmetry will bring about new viewpoints on superconducting mechanisms in BiCh$_2$-based superconductors.

Very recently, extremely high $B_{c2}$ was observed in LaO$_{0.5}$F$_{0.5}$BiS$_{2-x}$Se$_x$ ($x$ = 0.22, 0.69) single crystals [23]. In these compounds, the superconducting states were not destroyed even at a magnetic field strength of 55 T. This result implies that local inversion asymmetry plays a significant role in superconductivity, and the paramagnetic pair-breaking effect is suppressed by the Rashba-type ASOC. A high $B_{c2}$ value was also observed in LaO$_{0.5}$F$_{0.5}$BiS$_2$ and NdO$_{0.7}$F$_{0.3}$BiS$_2$ single crystals [24,25]. However, detailed measurements of $B_{c2}$ have not yet been conducted for Ce- and Pr-containing systems. Therefore, the investigation of $B_{c2}$ for these crystals is desired to further understand the effects of Rashba-type ASOC in REOBiCh$_2$-type superconductors.

The target materials of this study were CeOBiS$_{1.7}$Se$_{0.3}$ and PrO$_{0.5}$F$_{0.5}$BiS$_{2-x}$Se$_x$ ($x$ = 0, 0.3). In CeOBiS$_{1.7}$Se$_{0.3}$, bulk superconductivity was observed in single crystals and polycrystalline samples, with a $T_c$ of ~3 K [26,27]. Notably, external elemental substitution on the REO layer is not required to generate electron carriers in the BiCh$_2$ layers because the carriers are self-doped via the mixed-valence states of Ce [26,28]. In addition, a previous study indicated that superconductivity and ferromagnetism coexist in CeO$_{1-x}$F$_x$BiS$_2$ systems [29,30]; thus, we can expect novel behaviors related to magnetic ordering through $B_{c2}$ measurements for the Ce system. In contrast, bulk superconductivity was observed in polycrystalline samples PrO$_{1-x}$F$_x$BiS$_2$ ($x$ = 0.3–0.9) [31]. In the PrO$_{0.7}$F$_{0.3}$BiS$_2$ single crystal, the $T_c$ was 3.8 K, and a high superconducting anisotropy parameter of $\gamma$ = 53–58 was observed [32]. Because Se-substituted samples have not been reported for the Pr system, we synthesized single crystals of PrO$_{0.5}$F$_{0.5}$BiS$_{2-x}$Se$_x$ ($x$ = 0, 0.3) and investigated the effect of Se substitution on $B_{c2}$ in this study. Pr ions, which have two $f$-electrons ($S$ = 1, $L$ = 5, and $J$ = 4 for the ground state multiplet), are non-Kramer ions, which suggests that the crystal field can produce non-magnetic singlet states. Hence, the results for $B_{c2}$ may show clear differences between the Ce and Pr systems. As a result of this study, we show that the magnetic field-temperature phase diagrams for superconducting states in Ce-based and Pr-based systems exhibit clear differences.

**II. Experimental**

CeOBiS$_{1.7}$Se$_{0.3}$ and PrO$_{0.5}$F$_{0.5}$BiS$_{2-x}$Se$_x$ ($x$ = 0, 0.3) single crystals were grown using a high-temperature flux method in an evacuated quartz tube. First, polycrystalline samples with nominal compositions of CeOBiS$_{1.6}$Se$_{0.4}$ and PrO$_{0.5}$F$_{0.5}$BiS$_{2-x}$Se$_x$ ($x$ = 0, 0.4) were prepared by the solid-state reaction method, as described in Refs. 26 and 33. The obtained polycrystalline samples (~1.0 g) were mixed with CsCl flux at a molar ratio of 1:20 and then sealed in an evacuated quartz tube. The sealed tube was heated at 950 °C for 15 h and slowly cooled to 650 °C at a rate of −1.0 °C/h. After the furnace cooled to room temperature, the quartz tube was opened under an air atmosphere, and the CsCl flux in the samples was dissolved in pure water.

The obtained samples were characterized by single-crystal X-ray structural analysis with Mo-K$_\alpha$ ($\lambda$ = 0.71075 Å) radiation on an XtaLAB (Rigaku). The structural parameters were refined with the tetragonal (*P*4/*nmm*) structural model using the refinement program SHELXL [34]. A crystal structure image was depicted using VESTA software [35]. The single crystals were analyzed using scanning electron microscopy (SEM), and their chemical compositions were investigated using energy-dispersive X-ray spectroscopy (EDX), with the exception of light elements.

Electrical resistivity measurements were performed using a four-terminal method. The terminals were fabricated using Au wire and Ag paste. Resistivity measurements in magnetic fields of up to 9 T and 2 K were performed using a physical property measurement system (Quantum Design, PPMS) with a horizontal rotator. The temperature dependence of magnetic susceptibility was measured using a superconducting quantum interference device (SQUID) magnetometer with an applied field of 10 Oe after zero-field cooling (ZFC) and field cooling (FC) on an MPMS3 (Quantum Design).

**III. Results and discussion**

**A. Crystal structure**

Figure 1 shows the crystal structure of the target system REOBiCh$_2$. This system belongs to a centrosymmetric space group (*P*4/*nmm*) with a noncentrosymmetric polar site (point group $C_{4v}$ for Bi, Ch, and RE atoms) and noncentrosymmetric nonpolar sites (point group $S_4$ for O atoms). The crystal structure of REOBiCh$_2$ can be divided into three sectors: two inversion-partner BiCh$_2$ layers ($\alpha$-sector and $\beta$-sector in Fig.1) and the REO blocking layer [20]. The two adjacent BiCh$_2$ layers were weakly coupled by van der Waals forces. A theoretical study suggested that the amplitude of the ASOC needs to be larger than the interlayer coupling to affect the physical properties of locally noncentrosymmetric multilayer systems [18]. In BiCh$_2$-based compounds, the REO blocking layer does not affect the band structure near the Fermi surface and acts as a buffer layer. Therefore, the interlayer coupling is weak and the spin polarization does not cancel out between the $\alpha$-sector and $\beta$-sector.

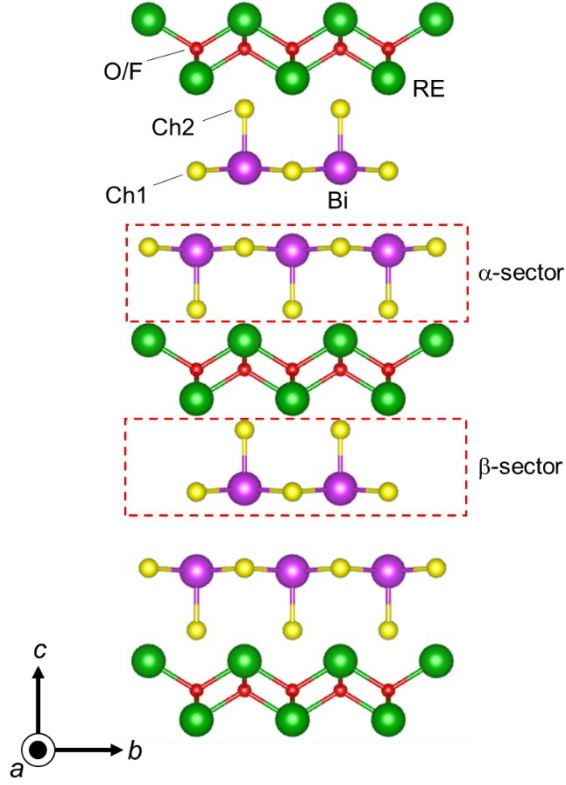

FIG. 1. Crystal structure image of REOBiCh$_2$ with a space group of $P4/nmm$ (No. 129). Ch1 and Ch2 denote the in-plane and out-of-plane chalcogen sites, respectively. The α-sector and β-sector represent the inversion BiCh$_2$ layers, and the inversion symmetry is broken in each sector.

We performed single-crystal X-ray structural analysis on PrO$_{0.5}$F$_{0.5}$BiS$_{2-x}$Se$_x$ ($x$ = 0, 0.3) single crystals. Details of the analysis conditions and the refined structural parameters of CeOBiS$_{1.7}$Se$_{0.3}$ and PrO$_{0.5}$F$_{0.5}$BiS$_{2-x}$Se$_x$ ($x$ = 0, 0.3) are summarized in Table I. CeOBiS$_{1.7}$Se$_{0.3}$ were taken from Ref. 27. Because Se ions selectively occupy the in-plane Ch1 sites [36], the Se occupation ratio for the Ch2 sites was fixed at 0. In PrO$_{0.5}$F$_{0.5}$BiS$_{2-x}$Se$_x$, the lattice parameters ($a$ and $c$) increase with Se substitution, which can be explained by the difference in the ionic radii of Se$^{2-}$ and S$^{2-}$. Similar elongations of lattice parameters by Se substitution have been observed in LaO$_{0.5}$F$_{0.5}$BiS$_{2-x}$Se$_x$ and CeOBiS$_{2-x}$Se$_x$ [26,37]. The average ratios of the constituent elements (except for O and F) from the EDX analysis are shown in Table II, where the RE value was fixed at 1 for normalization. The chemical ratios estimated using EDX were consistent with the site occupancies obtained from the X-ray structural analysis.

Table I. Refined crystal structure parameters of CeOBiS$_{1.7}$Se$_{0.3}$ and PrO$_{0.5}$F$_{0.5}$BiS$_{2-x}$Se$_x$ ($x$ = 0, 0.3). The data of CeOBiS$_{1.7}$Se$_{0.3}$ were taken from Ref. 27. The atomic coordinates are RE(0, 0.5 $z$), O/F(0, 0, 0), Bi(0, 0.5, $z$), Ch1(0, 0.5, $z$), and Ch2(0, 0.5, $z$).

|  | CeOBiS$_{1.7}$Se$_{0.3}$ | PrO$_{0.5}$F$_{0.5}$BiS$_2$ | PrO$_{0.5}$F$_{0.5}$BiS$_{1.7}$Se$_{0.3}$ |
|---|---|---|---|
| Formula weight | 443.29 | 431.51 | 445.58 |
| Space group | $P4/nmm$ (#129) | $P4/nmm$ (#129) | $P4/nmm$ (#129) |
| $a$ (Å) | 4.0327(9) | 4.0180(9) | 4.0348(9) |
| $c$ (Å) | 13.603(4) | 13.403(5) | 13.410(4) |
| $V$ (Å$^3$) | 221.22(9) | 216.38(9) | 218.31(9) |
| $z$ (RE) | 0.0915(2) | 0.0972(2) | 0.0966(1) |
| $z$ (Bi) | 0.6288(1) | 0.6256(1) | 0.6246(1) |
| $z$ (Ch1) | 0.3804(8) | 0.3767(9) | 0.3759(4) |
| $z$ (Ch2) | 0.8135(8) | 0.8120(7) | 0.8108(4) |
| Se occupancy at Ch1 | 0.28(5) | - | 0.31(4) |
| Se occupancy at Ch2 | 0 (fixed) | - | 0 (fixed) |
| Residuals: $R$ (All reflections) | 0.0793 | 0.0550 | 0.0408 |
| Goodness of fit indicator | 1.184 | 1.120 | 0.781 |
| Temperature (K) | 293 | 293 | 293 |

Table II. The average ratio of the constituent elements (except for O and F) from EDX analysis of CeOBiS$_{1.7}$Se$_{0.3}$ and PrO$_{0.5}$F$_{0.5}$BiS$_{2-x}$Se$_x$ ($x$ = 0, 0.3). The data of CeOBiS$_{1.7}$Se$_{0.3}$ were taken from Ref. 27. The RE value was fixed to 1 for normalization.

|  | CeOBiS$_{1.7}$Se$_{0.3}$ | PrO$_{0.5}$F$_{0.5}$BiS$_2$ | PrO$_{0.5}$F$_{0.5}$BiS$_{1.7}$Se$_{0.3}$ |
|---|---|---|---|
| RE | 1(fixed) | 1(fixed) | 1(fixed) |
| Bi | 1.00(1) | 0.97(2) | 1.01(2) |
| S | 1.74(2) | 2.06(1) | 1.83(1) |
| Se | 0.30(1) | - | 0.28(4) |

**B. CeOBiS$_{1.7}$Se$_{0.3}$**

Figure 2 shows the temperature dependence of (a) the $ab$-plane resistivity ($\rho_{ab}$) and (b) the magnetization after ZFC and FC with an applied field of 10 Oe parallel to the $c$-axis. The zero-resistivity temperature ($T_c^{\text{zero}}$) and $T_c$ estimated from magnetization measurements were ~3.3 K, which is almost the same as the previous result [27]. In addition, a large diamagnetic signal corresponding to superconductivity was observed, indicating that the observed superconducting states were bulk in nature.

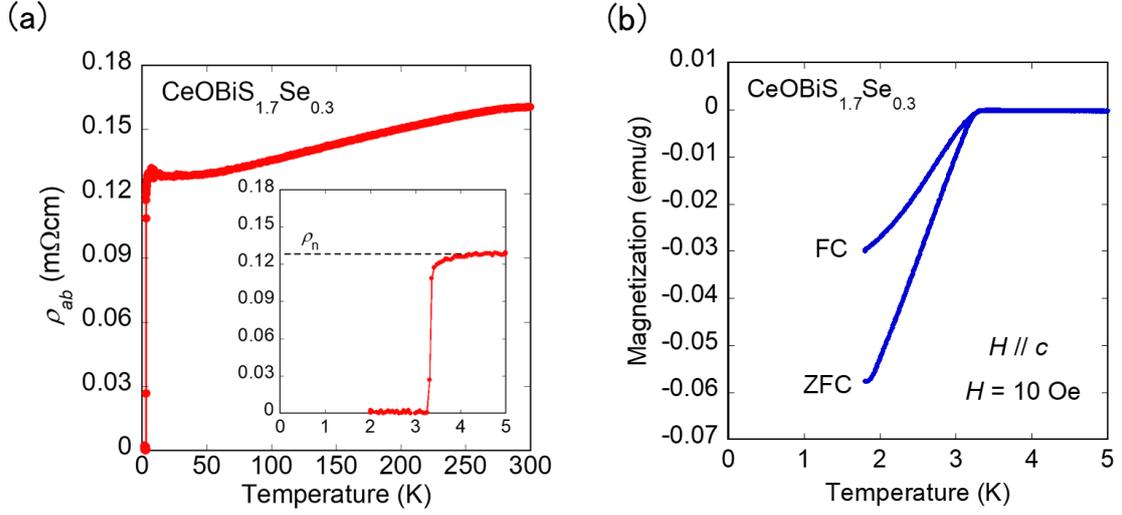

FIG. 2. Temperature dependence of (a) the *ab*-plane resistivity $\rho_{ab}$ and (b) the magnetization (ZFC and FC). The black dashed line denotes the normal state resistivity $\rho_n$ just above $T_c$.

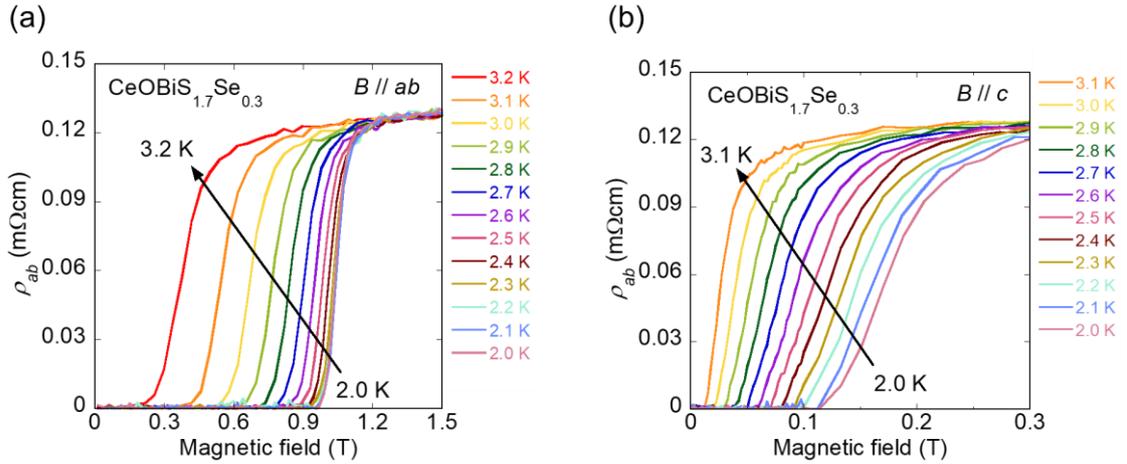

FIG. 3. The field dependence of the *ab*-plane resistivity $\rho_{ab}$ at different temperatures with (a) *B//ab* and (b) *B//c*. The arrow indicates the direction from low to high temperatures.

Figures 3(a) and 3(b) show the field dependence of $\rho_{ab}$ at different temperatures for *B//ab* and *B//c*, respectively. In BiCh$_2$-based superconductors, the robustness of the superconducting state against the *ab*-plane magnetic field has been reported, and the superconducting states survive under *ab*-plane magnetic fields higher than 9 T [23-25]. In contrast, the superconducting states in CeOBiS$_{1.7}$Se$_{0.3}$ are suppressed under the *ab*-plane magnetic field below ~1.2 T (Fig.3(a)). Interestingly, even though the temperature decreased from 2.5 K to 2 K, the superconducting states were destroyed under almost the same magnetic field. In the case of $B//c$, the superconducting states were immediately suppressed by the applied field, similar to other BiCh$_2$-based superconductors [23-25]. The behaviors observed in Fig. 3(a) and 3(b) are consistent with the temperature dependence of $\rho_{ab}$ under various magnetic fields with $B//ab$ and $B//c$ (Fig. S1 in the Supplemental Material [38]).

Figure 4 shows $B_{c2}$ with (a) $B//ab$ and (b) $B//c$ as functions of the (normalized) temperature $t = T/T_c$. We defined three characteristic fields: the high-field onset of the superconducting transition (90% of $\rho_n$), mid-point field (50% of $\rho_n$), and zero-resistivity field (0% of $\rho_n$). Here, we define the normal state resistivity ($\rho_n$) as the black dashed line in Fig. 2(a). We estimated the orbital limit $B_{c2}^{orb}(0)$ from the initial slope of $B_{c2}$ at $T_c$ based on the relation $B_{c2}^{orb}(0) = 0.69T_c(-dB_{c2}/dT)_{Tc}$ in the dirty limit [39] and describe the Werthamer–Helfand–Hohenberg (WHH) curves (dashed lines in Fig. 4). In the case of $B//ab$, we found a clear difference between the observed data and WHH curves, suggesting that the orbital pair-breaking effect is not dominant. Furthermore, the value of $B_{c2}^{//ab}$ gradually approached a constant value in the low-temperature region. However, the orbital pair-breaking effect is dominant in the case of $B//c$, which is confirmed by the consistency between the WHH curves and the experimental data. For the paramagnetic pair-breaking effect, the Pauli limit is estimated from the relation $B_P(0) = 1.86T_c$ if we assume spin-singlet and weak-coupling superconductors. The Pauli limit is estimated to be 6.1 T ( black diamond in Fig. 4(a)), and this value is much higher than the observed $B_{c2}^{//ab}$. The superconducting anisotropic parameter $\gamma$ was estimated to be 5.9 at 2.0 K based on the relation $\gamma = B_{c2}^{//ab}/ B_{c2}^{//c}$. This value is much lower than those of other BiCh$_2$-based superconductors [40] because of the smallness of $B_{c2}^{//ab}$.

Herein, we discuss the origin of this unexpectedly low $B_{c2}^{//ab}$. A previous study of electrical resistivity, magnetization, and specific heat measurements in CeOBiS$_2$ single crystals showed that the Ce 4$f$-electron state is well-localized split by crystalline-electric-field (CEF) effects, and the CEF ground state is a pure $|\pm 1/2\rangle$ doublet [41]. In addition, the $ab$-plane was found to be a magnetic easy plane. In neutron scattering measurements of high-pressure-annealed CeO$_{0.7}$F$_{0.3}$BiS$_2$ polycrystals with superconductivity, the magnetic moments were found to be ferromagnetically aligned along the $c$-axis below 7 K, and the ferromagnetic component developed in the $ab$-plane by applying an external magnetic field [30]. A similar magnetic structure was observed for CeAgSb$_2$ [42]. Given these previous studies, spins possibly flop to the $ab$-plane by applying a magnetic field of ~ 1 T, which forms ferromagnetic ordering along the $ab$-plane in CeOBiS$_{1.7}$Se$_{0.3}$. Consequently, the internal magnetic field due to ferromagnetic ordering may break the superconductivity.

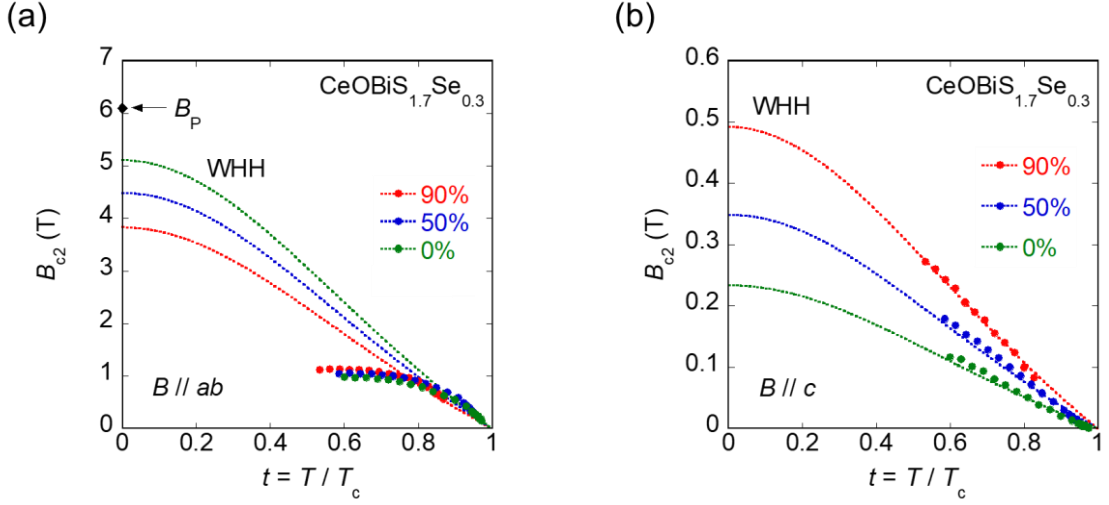

FIG. 4. The upper critical field $B_{c2}$ with (a) $B//ab$ and (b) $B//c$ as a function of (normalized) temperature $t = T/T_c$. The $B_{c2}$ is defined by the high-field onset of superconducting transition (90% of $\rho_n$), the midpoint field (50% of $\rho_n$), and the zero-resistivity field (0% of $\rho_n$). The dashed curves show the Werthamer–Helfand–Hohenberg (WHH) fits. The black diamond denotes the Pauli limit.

Figure 5(a) shows a schematic of the rotation angle $\theta$ for upper critical field measurements. The angular dependence of $B_{c2}$ at 2 K is shown in Figure 5(b). The value of $B_{c2}(\theta)$ was estimated from the midpoint of the resistivity transition. The solid and dashed lines in Fig. 5(b) represent the anisotropic three-dimensional (3D) Ginzburg-Landau (GL) model and the two-dimensional (2D) Tinkham model, respectively. In the anisotropic-3D-GL model, the angular $\theta$ dependence of $B_{c2}$ is expressed as the relation $(B_{c2}(\theta)\cos\theta/B_{c2}^{//c})^2 + (B_{c2}(\theta)\sin\theta/B_{c2}^{//ab})^2 = 1$. In contrast, in the 2D-Tinkham model, the angular $\theta$ dependence of $B_{c2}$ is expressed as $|B_{c2}(\theta)\cos\theta/B_{c2}^{//c}| + (B_{c2}(\theta)\sin\theta/B_{c2}^{//ab})^2 = 1$. The experimental data were well described by the 2D-Tinkham model, suggesting that two-dimensional nature is dominant in CeOBiS$_{1.7}$Se$_{0.3}$. This result supports the scenario of ferromagnetic ordering along the *ab*-plane. At 3 K, the two dimensionalities were slightly lower (Fig. S2 in the Supplemental Material [38]).

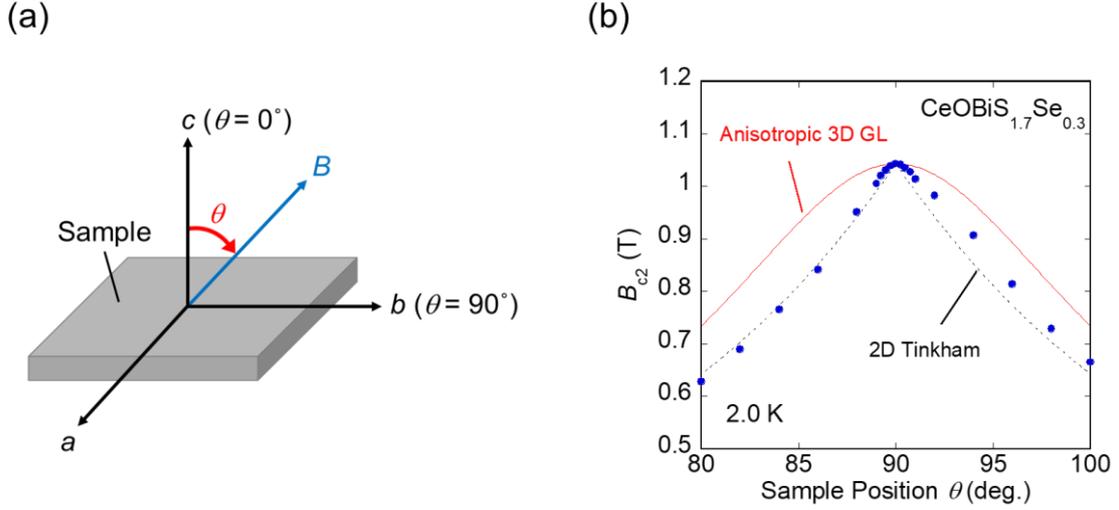

FIG. 5. (a) Schematic image of the rotation angle. (b) The angular dependence of the upper critical field $B_{c2}$ at 2.0 K. The solid line denotes the anisotropic 3D-Ginzburg-Landau (GL) model. The dashed line denotes the 2D-Tinkham model.

### C. $PrO_{0.5}F_{0.5}BiS_{2-x}Se_x$

Figures 6(a) and 6(b) show the temperature dependences of $\rho_{ab}$ for $x = 0$ and 0.3, respectively. The $T_c^{zero}$ in both samples was ~4.5 K, and the $T_c^{zero}$ for $x = 0$ was consistent with that observed in a previous study [32]. In the normal state, a broad anomaly was observed at approximately 100 K for $x = 0$. Similar behaviors have been observed in $PrO_{0.7}F_{0.3}BiS_2$ and $NdO_{1-x}F_xBiS_2$ single crystals [32,43]. Although CDW-like instability has been pointed out for $BiCh_2$-based compounds [44-46], the origin of the observed anomaly is unclear at present. Se substitution suppresses the anomalous behavior, as shown in Fig. 6(b). Figure 6(c) shows the temperature dependence of the magnetization after ZFC and FC with an applied field of 10 Oe parallel to the $c$-axis. We can confirm the bulk nature of the superconductivity. The irreversible temperature $T_c^{irr}$, which is a bifurcation point between the ZFC and FC curves, is 4.4 K and 4.5 K for $x = 0$ and 0.3, respectively. In this system, Se substitution did not significantly affect the $T_c$. The phase diagram in $CeO_{0.5}F_{0.5}BiS_{2-x}Se_x$ exhibits dome-shaped behavior upon Se substitution, while the Se substitution in $NdO_{0.5}F_{0.5}BiS_{2-x}Se_x$ decreases $T_c$ [47]. Therefore, applying too much CP negatively affects superconductivity in $NdO_{0.5}F_{0.5}BiS_{2-x}Se_x$. Our results imply that the $PrO_{0.5}F_{0.5}BiS_{2-x}Se_x$ system lies in the middle region between the Ce and Nd systems.

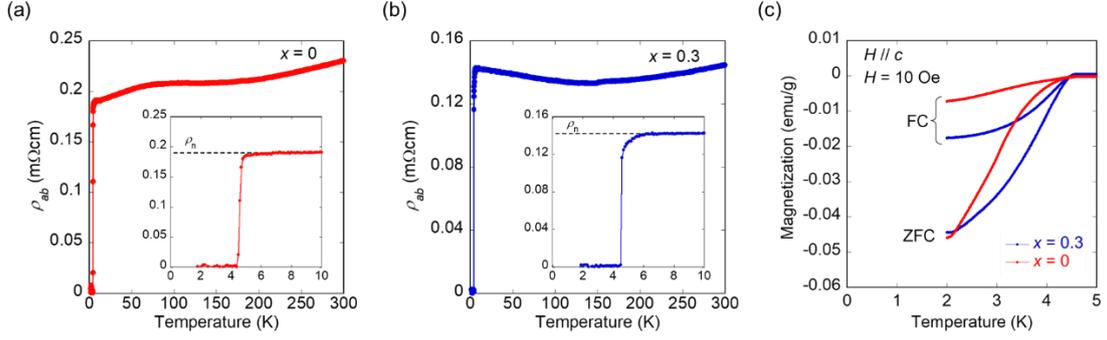

FIG. 6. The temperature dependence of the *ab*-plane resistivity ($\rho_{ab}$) for (a) $x = 0$ and (b) $x = 0.3$. The black dashed lines denote the normal state resistivity ($\rho_n$) just above $T_c$. (c) The temperature dependence of the magnetization (ZFC and FC) for $x = 0$ and 0.3.

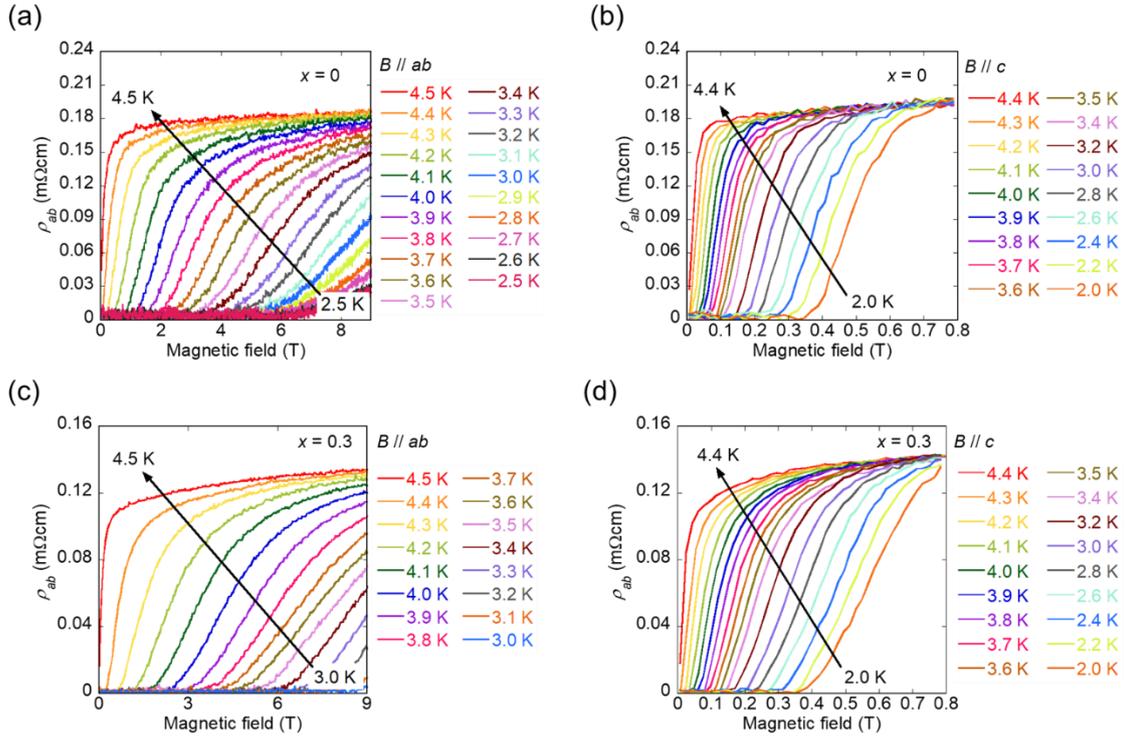

FIG. 7. The field dependence of the *ab*-plane resistivity $\rho_{ab}$ at different temperatures for $x = 0$ with (a) *B//ab* and (b) *B//c* and for $x = 0.3$ with (c) *B//ab* and (d) *B//c*. The arrow indicates the direction from low to high temperatures.

Figure 7 shows the field dependence of $\rho_{ab}$ for $x = 0$ at different temperatures with (a) *B//ab* and (b) *B//c*, and for $x = 0.3$ at different temperatures with (c) *B//ab* and (d) *B//c*. In the case of *B//ab*, the superconductivity was robust against the applied field in both samples. By contrast, in the case of $B//c$, the superconducting states were immediately suppressed by the field. The robustness of superconductivity under the fields for $x = 0.3$ is remarkable in the case of $x = 0$. The behaviors in Fig. 7 are consistent with the temperature dependence of $\rho_{ab}$ under various magnetic fields (Fig. S3 in the Supplemental Material [38]).

Figure 8 shows $B_{c2}$ as a function of (normalized) temperature $t = T/T_c$ for $x = 0$ with (a) $B//ab$ and (b) $B//c$ and for $x = 0.3$ with (c) $B//ab$ and (d) $B//c$. The definitions of the three types of upper critical field are the same as those in Fig. 4. We define $\rho_n$ as the black dashed line in Fig. 6(a) and 6(b). In the case of $B//ab$, the orbital pair-breaking effect was dominant in both samples because the WHH curves described the experimental data well. In general, the orbital limit is enhanced when the coherence length along the $c$-axis ($\xi_c$) is shorter than the thickness of the REO blocking layer, and a Josephson vortex state is formed. In the Josephson vortex state, the orbital limit deviates from the WHH curve. We estimated $\xi_c$ from the relations $B_{c2}^{//c} = \Phi_0/2\pi\xi_{ab}^2$ and $B_{c2}^{//ab} = \Phi_0/2\pi\xi_{ab}\xi_c$. This led to 11 Å for $x = 0$ at 3.2 K, and 10 Å for $x = 0.3$ at 3.5 K. These values are larger than the REO blocking layer thickness of 2.6 Å (293 K). Therefore, the experimental data fitted the WHH curves. In contrast, the paramagnetic pair-breaking effect was suppressed in both samples. The Pauli limit is estimated to be 8.4 T (the black diamonds in Figs. 8(a) and 8(b)) from the relation $B_P(0) = 1.86T_c$. The experimental data exceeded this value at a finite temperature, and possibly exhibited a higher upper critical field at 0 K. Some experimental results for BiCh$_2$-based superconductors indicate a strong coupling nature [48-50], which leads to the enhancement of the Pauli limit. However, even if we use the reported value of $2\Delta/k_BT_c = 3.4$–4.5, the estimated Pauli limit is smaller than 11 T. Another possibility is that the two-gap nature was reported in muon spin-relaxation measurements on LaO$_{0.5}$F$_{0.5}$BiS$_2$ [51] and scanning tunneling spectroscopy on NdO$_{0.5}$F$_{0.5}$BiS$_2$ [52]. In the LaO$_{0.5}$F$_{0.5}$BiS$_2$ single crystal, the two-gap model explained the large upper critical field well [24]. However, the specific heat measurement of LaO$_{0.5}$F$_{0.5}$BiSSe single crystals indicates a single-gap nature [50]. We suggested that local inversion asymmetry enhances the Pauli limit in LaO$_{0.5}$F$_{0.5}$BiS$_{2-x}$Se$_x$ ($x = 0.22, 0.69$) single crystals [23]. In tetragonal systems with the Rashba-type ASOC, some components of magnetic susceptibility do not change after the superconducting transition, resulting in protection of the Cooper pairs from de-pairing against the applied fields [18,24,53]. To confirm the effect of Rashba-type ASOC on PrO$_{0.5}$F$_{0.5}$BiS$_{2-x}$Se$_x$, upper critical field measurements below 3 K are required.

In the case of $B//c$, the paramagnetic pair-breaking effect is suppressed in local inversion asymmetry systems with a tetragonal structure [17,18]. Thus, the orbital pair-breaking effect was dominant in this system. However, the experimental data derived from the WHH curves in the low-temperature region are shown in Fig. 8(b) and 8(d). A similar behavior was observed for LaO$_{0.5}$F$_{0.5}$BiS$_{2-x}$Se$_x$ [23,50]. Although the origin of this behavior is currently unclear, the two-gap model fits well with the experimental data for LaO$_{0.5}$F$_{0.5}$BiS$_2$ [24]. The superconducting anisotropic parameter ($\gamma$) was estimated to be 34.2 ($x = 0$) and 38.3 ($x = 0.4$) at 3.5 K.

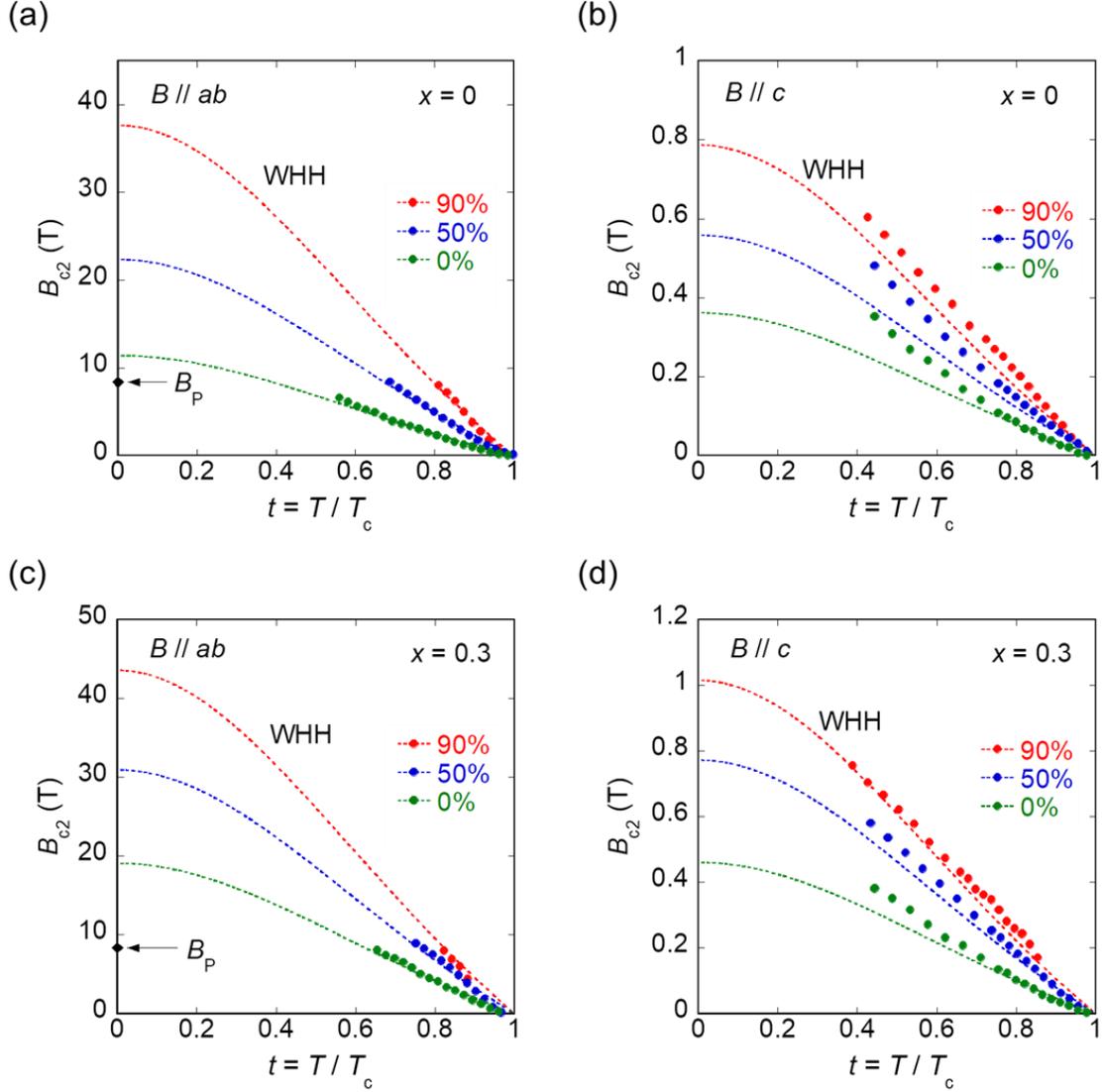

FIG. 8. The upper critical field ($B_{c2}$) as a function of (normalized) temperature $t = T/T_c$. The temperature dependence of $B_{c2}$ for $x = 0$ with (a) $B//ab$ and (b) $B//c$, and for $x = 0.3$ with (c) $B//ab$ and (d) $B//c$. $B_{c2}$ is defined by the high-field onset of superconducting transition (90% of $\rho_n$), the mid-point field (50% of $\rho_n$), and the zero-resistivity field (0% of $\rho_n$). The dashed curves show the Werthamer–Helfand–Hohenberg (WHH) fits. The black diamond denotes the Pauli limit.

Figure 9 shows the angular dependence of $B_{c2}$ for (a) $x = 0$ at 3.3 K and (b) $x = 0.3$ at 3.7 K. The definitions of $B_{c2}(\theta)$, the solid line, and the dashed line are the same as those in Fig. 5. The experimental data are described by the anisotropic 3D-GL model for $x = 0$, suggesting that the 3D nature is dominant. In contrast, the two dimensionalities were enhanced for $x = 0.4$, whereas the experimental data did not fit perfectly with the 2D-Tinkham model. These results indicate that the Abrikosov vortex states or the crossover of the conventional Abrikosov and Josephson vortex states was realized in this system, which is consistent with the behavior of $B_{c2}^{//ab}$ shown in Figs. 8(a) and 8(c). The enhancement of two dimensionalities implies that the Rashba-type ASOC increases with Se substitution because the Rashba-type ASOC generally weakens the interlayer coupling and enhances the 2D nature of the superconductivity [18]. However, a theoretical study suggested that the amplitude of the ASOC is larger in $LaOBiS_2$ than in $LaOBiSe_2$ [22]. In addition, the thickness of a single crystal or the bulk nature of the samples may affect the behavior. Measurement of the angular dependence of the upper critical field using a thin film is highly desirable. The angular dependence of $B_{c2}$ at different temperatures is shown in Fig. S4 in the Supplemental Material [38].

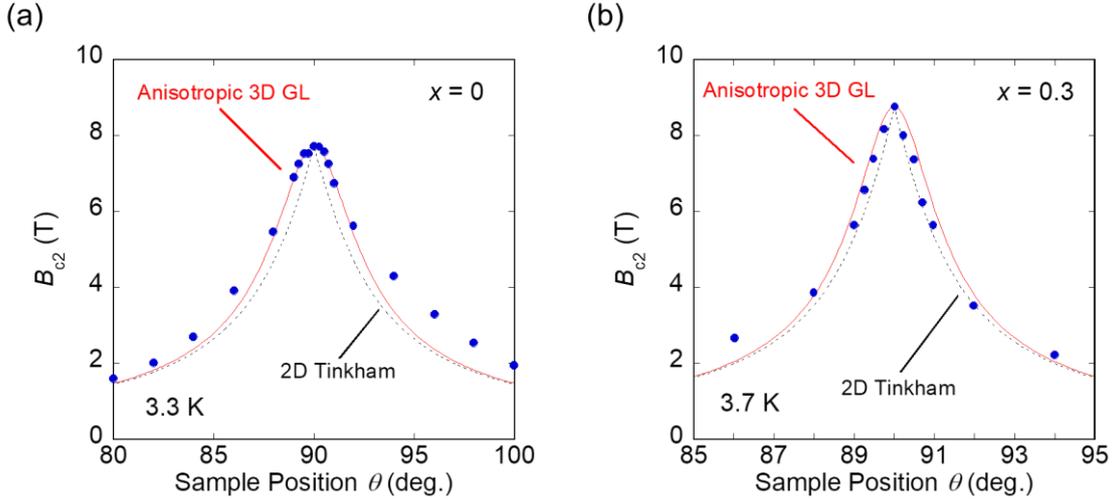

FIG. 9. The angular dependence of the upper critical field $B_{c2}$ for (a) $x = 0$ at 3.3 K and (b) $x = 0.3$ at 3.7 K. The solid line denotes the anisotropic 3D-Ginzburg-Landau (GL) model. The dashed line denotes the 2D-Tinkham model.

## IV. Conclusions

We investigated $B_{c2}$ of single crystals of $BiCh_2$-based layered superconductors $CeOBiS_{1.7}Se_{0.3}$ and $PrO_{0.5}F_{0.5}BiS_{2-x}Se_x$ ($x = 0, 0.3$). In $CeOBiS_{1.7}Se_{0.3}$, we observed that $B_{c2}^{//ab}$ was smaller than the conventional orbital and Pauli limits. This result implies that ferromagnetic ordering along the $ab$-plane affects $B_{c2}$ and superconductivity. In contrast, a high $B_{c2}^{//ab}$ ratio was observed for $PrO_{0.5}F_{0.5}BiS_{2-x}Se_x$ ($x = 0, 0.3$). The temperature dependence of $B_{c2}^{//ab}$ was well described by the WHH model, and at low temperatures, $B_{c2}^{//ab}$ exceeded the Pauli limit. The suppression of the paramagnetic pair-breaking effect is due to Rashba-type ASOC, which arises from the lack of local inversion symmetry. In REOBiCh$_2$-based compounds, local inversion asymmetry and the characteristics of the rare-earth element in the REO layer are essential for superconducting properties and pairing mechanisms.

ACKNOWLEDGMENTS

We thank O, Miura, Y. Aoki, and T. D. Matsuda for their support with the experiments and discussions. This work was partly supported by JSPS-KAKENHI (20J21627, 18KK0076, 21H00151) and Tokyo Metropolitan Government Advanced Research (H31-1). We would like to thank Editage (www.editage.com) for English language editing.

**Supplemental Material**

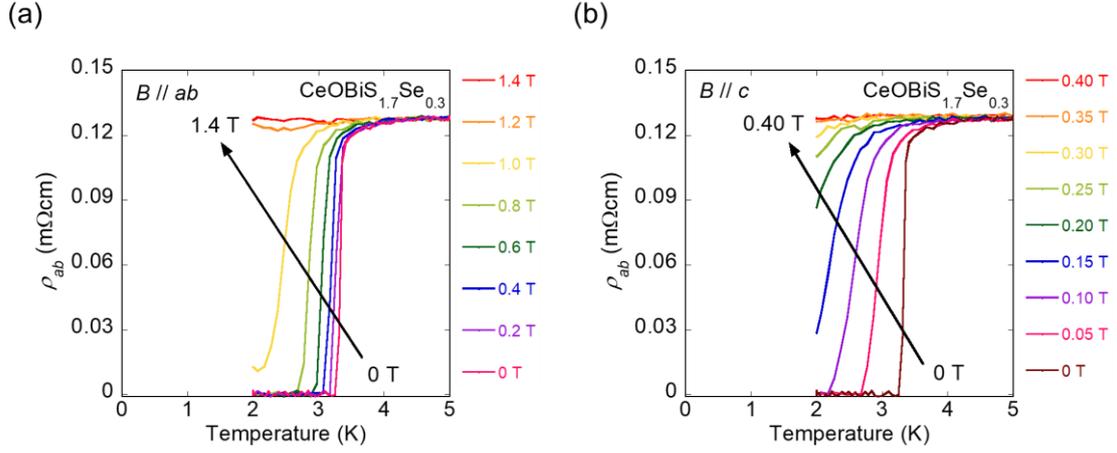

FIG. S1. Temperature dependence of the *ab*-plane resistivity ($\rho_{ab}$) for a CeOBiS$_{1.7}$Se$_{0.3}$ single crystals under various magnetic fields along (a) *B//ab* and (b) *B//c*. The arrow indicates the direction from low to high magnetic fields.

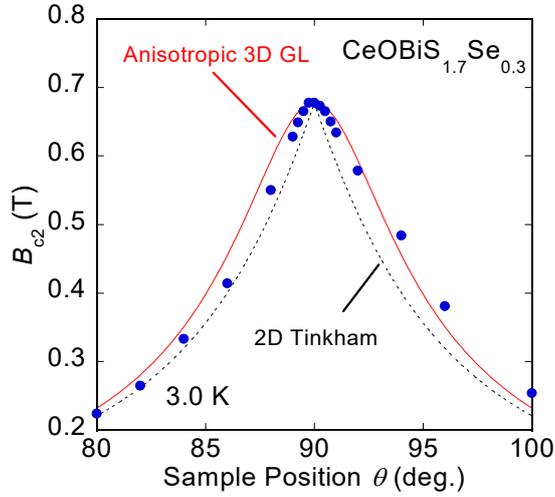

FIG. S2. Angular dependence of the upper critical field ($B_{c2}$) at 3.0 K for a CeOBiS$_{1.7}$Se$_{0.3}$ single crystal. The solid line denotes the anisotropic 3D-Ginzburg-Landau (GL) model. The dashed line denotes the 2D-Tinkham model.

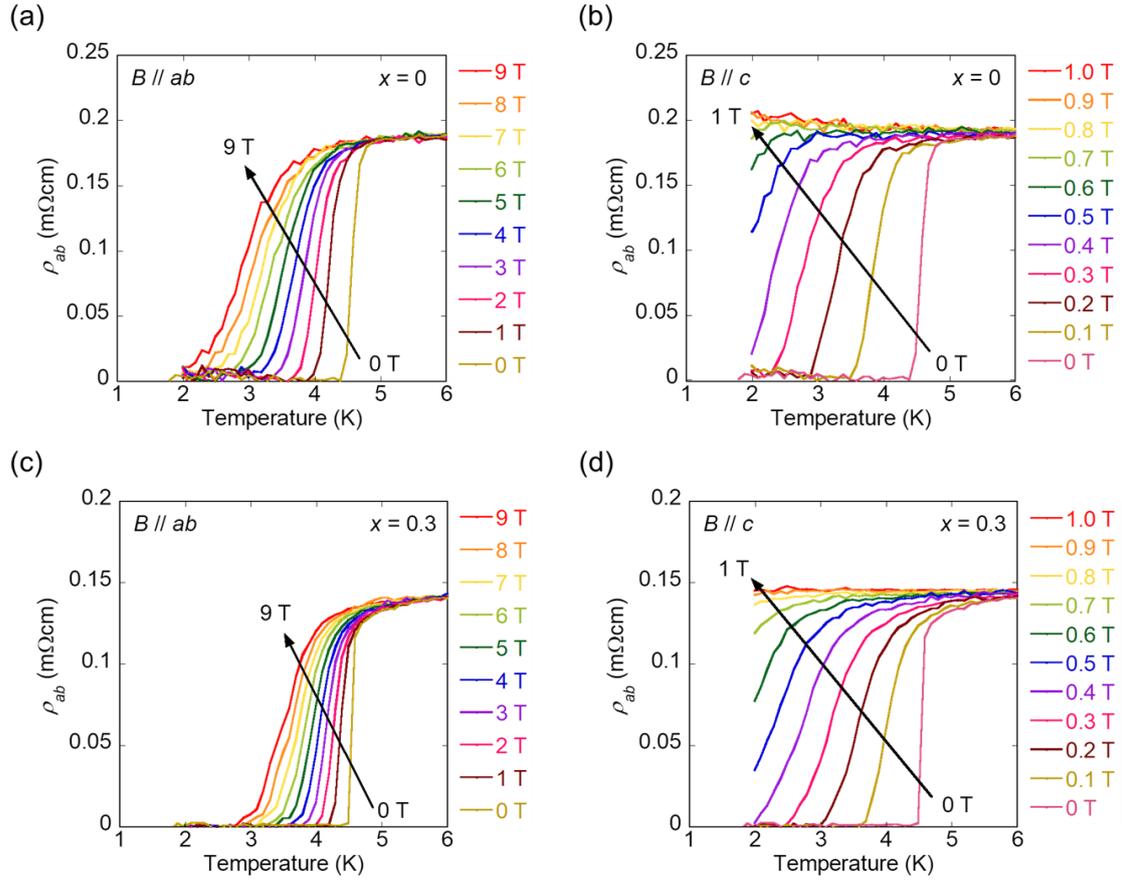

FIG. S3. Temperature dependence of the $ab$-plane resistivity ($\rho_{ab}$) for PrO$_{0.5}$F$_{0.5}$BiS$_{2-x}$Se$_x$ ($x$ = 0, 0.3) single crystals under various magnetic fields; for $x$ = 0 with (a) $B//ab$ and (b) $B//c$, and for $x$ = 0.3 with (c) $B//ab$ and (d) $B//c$. The arrow indicates the direction from low to high magnetic fields.

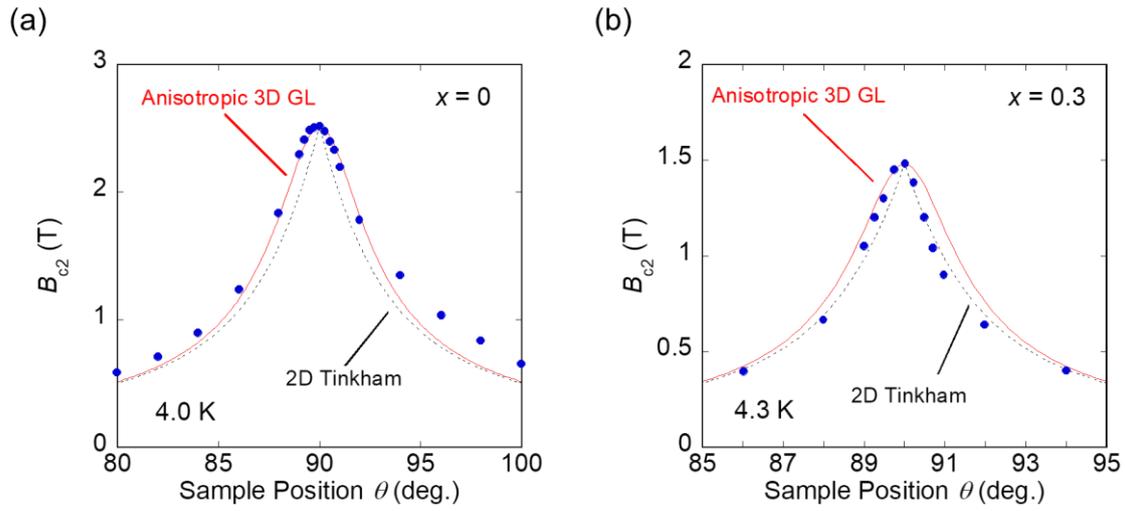

FIG. S4. Angular dependence of the upper critical field ($B_{c2}$) for PrO$_{0.5}$F$_{0.5}$BiS$_{2-x}$Se$_x$ ($x$ = 0, 0.3) single crystals; for (a) $x$ = 0 at 4.0 K and (b) $x$ = 0.3 at 4.3 K. The solid line denotes the anisotropic 3D-Ginzburg-Landau (GL) model. The dashed line denotes the 2D-Tinkham model.